\documentstyle[prb,aps,manuscript,twocolumn,epsf]{revtex}  
\begin{document}
\tighten
\title{Silence of magnetic layers to magnetoresistive process and
electronic separation at low temperatures in (La, Sm)Mn$_2$Ge$_2$}

\author{E.V. Sampathkumaran,$^*$ R. Mallik, P.L. Paulose and Subham
Majumdar}

\address{Tata Institute of Fundamental Research, Homi  Bhabha  Road
Mumbai-400005, INDIA}
\maketitle

\begin{abstract}
A closer look at the temperature (T) dependence of magnetoresistance (MR)
of two polycrystalline magnetic compounds, LaMn$_2$Ge$_2$ and
SmMn$_2$Ge$_2$, previously reported by us, is made. A common feature for
both these compounds is that the low temperature MR is positive (say,
below, 30 K) in spite of the fact that both are ferromagnetic at such low
temperatures; in addition, MR as a function of magnetic field (H) does not
track magnetization (M) in the sense that M saturates at low fields, while
MR varies linearly with H.  These observations suggest that the magnetic
layers interestingly do not dominate low temperature magnetotransport
process. Interestingly enough, as the T is increased, say around 100 K,
these magnetic layers dominate MR process as evidenced by the tracking of
M and MR in SmMn$_2$Ge$_2$. These results tempts us to propose that there
is an unusual "electronic separation" for MR process as the T is lowered
in this class of compounds.
\end{abstract}
\vskip0.6cm


\noindent $^*$E-mail: sampath@tifr.res.in

\newpage
The understanding of electrical resistance (R) behaviour in solids
continues to be a major direction of research in condensed matter physics.
In general, in metals containing magnetic impurities or a large
concentration of magnetic-moment-carrying ions, one observes a dominant
signature of these magnetic ions in the {\it low temperature} (T)
electrical resistance, which has a characteristic response with the
application of an external magnetic field (H)~\cite{Pippard} depending
upon the magnetic state. For instance, in ferromagnetic materials R
decreases with increasing H; Y (non-magnetic), if contains Ce in traces,
exhibits features due to the Kondo effect resulting in a  decrease of R
with H~\cite{Maple}. Here we argue that, in sharp contrast to this
expectation, in a class of layered ternary intermetallic compounds
containing a magnetically ordered layer, viz., RMn$_2$Ge$_2$ (R= La, Sm),
the non-magnetic atomic layers  dominate the magnetoresistance (MR) at low
temperatures with the magnetic layers apparently remaining "silent";
however, with increasing temperature, the contribution from the magnetic
atomic layers is visible.  This is taken as an evidence to suggest that
there is a temperature dependence associated for the involvement of
different layers; in other words apparently there is "an electronic
separation" with decreasing temperature for the magnetoresistive process.
This finding adds a new dimension to the understanding of magnetotransport
phenomena in solids.
 
The intermetallic compounds under discussion crystallize  in the
well-known ThCr$_2$Si$_2$-type tetragonal structure~\cite{Szytula},
containing layers of atoms stacked in the sequence Th-Si-Cr-Si-Th along
the c-axis. It is interesting to note that, among few hundred compounds
known to form in this structure, the Mn is the only transition metal ion
known to possess magnetic moment and magnetic ordering at very high
temperatures. The nature of the magnetic ordering apparently is sensitive
to Mn-Mn distances and this appears to explain the presence of multiple
magnetic transitions for Sm compound with the Mn-Mn distance for this
compound falling near the critical limit [3]. Thus, while the former
compound has been known to order ferromagnetically at about 300 K with the
magnetism arising from Mn sublattice, the latter exhibits multiple
magnetic transitions: below about 345 K, para-  to ferro-magnetism; at
about 140 K, ferro to anti-ferromagnetism; and antiferro to ferromagnetism
below about 105 K (see, for instance, Refs. 3-13). It is surprising to
note that, inspite of extensive magnetic investigations on this class of
ternary compounds, virtually there has been very little MR studies on
these Mn alloys. Though the compound SmMn$_2$Ge$_2$ was investigated in
the range 80 - 150 K to show giant magnetoresistance effects
earlier~\cite{Dover,Brabers}, surprisingly these authors did not report
the MR at lower temperatures.  Considering a recent upsurge in MR studies
in condensed matter physics, we considered it important to carry out such
studies on these compounds and thus we reported the MR alongwith detailed
magnetic measurements down to 4.2
K~\cite{Sampath,Mallik1,Mallik2,Majumdar}; the major point of emphasis was
to bring out the relevance of these compounds to the physics of artificial
multilayers. In this letter, for  the La and Sm compounds, we take into
account available data  in the literature and  also compare   the T
dependence of MR to draw the present conclusion. As far as our data is
concerned, we reproduce here only those data which are required to
emphasize this point.
\begin{figure}
\epsfxsize = 7cm
\epsfbox{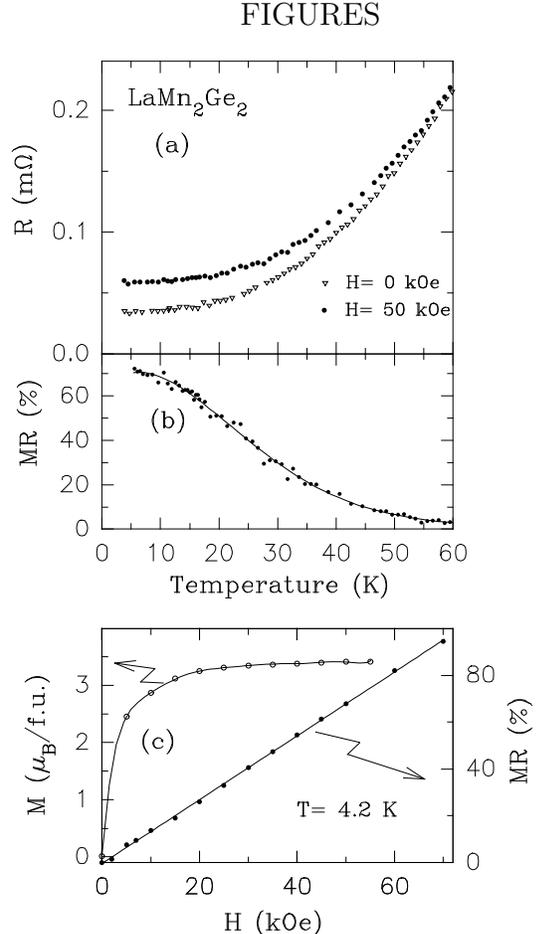}
\caption{ Electrical resistance (R) and magnetization behaviour of
LaMn$_2$Ge$_2$: {\bf(a)} The R data in zero field and in the presence of
50 kOe; {\bf(b)} The derived magnetoresistance (MR= R(H)-R(0)/R(0))  as a
function of temperature (4.2-60 K). The typical isothermal magnetization
data and MR  as a function of externally applied magnetic field in the
ferromagnetically ordered state, say at 4.2 K, are shown in {\bf(c)}. The
lines through the data points are guides to the eyes.}

\end{figure}

In figure 1a, we show R as a function of T in zero field as well as at 50
kOe (with the direction of H being parallel to that of the excitation
current) for a specimen of LaMn$_2$Ge$_2$ below 60 K.  The
magnetoresistance [MR=\{R(H)-R(0)\}/R(0)] behaviour, obtained by
subtracting the zero-field R data from that at 50 kOe is shown in figure
1b. The value of MR increases with decreasing temperature below 60 K
eventually attaining large values at 4.2 K; at temperatures higher than 60
K the value of MR is negligibly small. {\it What is intriguing is that the
sign of MR is positive at low temperatures}, inspite of the fact that the
compound contains ferromagnetic Mn layers in which case one should have
observed MR with a negative sign expected for polycrystalline
ferromagnetic metals~\cite{Pippard}.  (Therefore, it is the ferromagnetic
nature of the magnetic ordering that enables us to draw the present
conclusions on firm grounds, as positive MR for antiferromagnetics is
usually expected.)  We have obtained  MR on several specimens of this
compound prepared under different conditions of heat-treatment and the
positive sign of MR is always reproducable~\cite{Mallik2,Majumdar}.The 
 
\begin{figure}
\epsfxsize = 7cm
\epsfbox{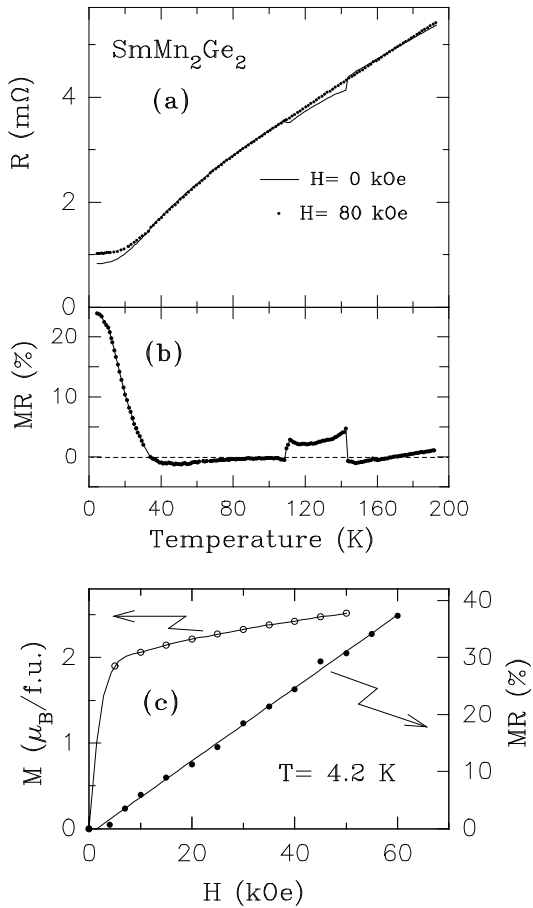}

\caption{Electrical resistance (R) and magnetization behaviour of
SmMn$_2$Ge$_2$: {\bf(a)} The R in the presence and in the absence of
magnetic field (50 kOe) as a function of temperature. The
magnetoresistance derived from this data are plotted in {\bf(b)}. The
isothermal magnetization data and MR  as a function of externally applied
magnetic field at 4.2 K are shown in {\bf(c)} and the lines through the
data points are guides to the eyes.}
\end{figure}
\noindent positive sign of MR  is taken as an evidence for the dominant role of
non-magnetic layers in controlling magnetoresistive process. Needless to
mention that the positive sign of MR is characteristic of non-magnetic
metals. Even if one attributes positive sign of MR to possible
deviations~\cite{Venturini} from collinear ferromagnetism of the Mn
sublattice or to any other mechanism arising from ferromagnetism, the
following finding goes against these possibilities. That is, the
magnetization does not track MR (e.g., at 4.2 K, see Fig. 1c) in the sense
that  M saturates for small applications of H while MR is a linearly
varying function of H. This is a {\it key finding} in favour of our
proposal that in this compound the Mn magnetic layer is not apparently
involved in the magnetotransport phenomena  at low temperatures. In
metallic materials, at low temperatures, the disorder/impurity scattering
contribution dominates as the phonon contribution tends to vanish.
Therefore, one may advance an arguement that  the positive sign of MR
somehow arises from such crystallographic imperfections. However, a
careful look at how MR varies as a function of the residual resistivity
ratio, RRR [= R(4.2K)/R(300K)], in specimens with different degree of
crystallographic imperfections~\cite{Majumdar} clearly reveals that the
increasing disorder/imperfections in fact tend to {\it diminish} the net
change in R by the application of H at low T.  This naturally implies that
the observed positive sign with a large magnitude is intrinsic to the
crystallographically well-ordered material. We therefore conclude that the
positive MR arises from La and/or Ge layer only, without involving
magnetic-moment-carrying Mn ions.

  The present conclusion is novel as it
has been generally believed that the signature of moment-carrying ions
should be prominent in the low temperature transport behaviour; this
contribution can be  so prominent that even in the dilute limit of 3d
magnetic impurities in non-magnetic matrices one usually encounters the
phenomenon of the Kondo effect~\cite{Maple,Grewe} with corresponding
magnetoresistive response. [It may be added that the temperature
dependence of mean free path (mfp) also seems to correlate with
MR~\cite{Majumdar}. Inspite of the fact that there are uncertainties in
the absolute values of mfp due to  approximations in its determination, it
appears that the critical value for the observation of significant
positive MR is of the order of interlayer spacings. We believe that this
information may be useful for future theoretical work.]

Next, one might be tempted to ask the following questions. Which of the
two layers, La or Ge, dominates low temperature scattering phenomenon? Do
magnetic layers participate in magnetotransport phenomena as the
temperature is increased? The transport behaviour of SmMn$_2$Ge$_2$ offers
straightforward answers for these questions. The data are shown in Fig. 2a
in zero and 50 kOe field. MR is distinctly positive at low temperatures
even in this compound, without tracking isothermal M (see Fig. 2c), the
origin of which must be the same as that in the La
compound.~\cite{Szytula} However, as the temperature is increased, as
known earlier~\cite{Szytula}, the jumps in R in zero field at about  110 K
and 140 K attributable to magnetic transitions arising from Mn are clearly
seen. The application of H wipes out these jumps due to the
well-known~\cite{Szytula,Dover} sharp metamagnetic transition occuring at
about 5 kOe, thus resulting in large MR anomalies at these temperatures
(Fig. 2b). Thus, it is evident that the Mn (magnetic) layer gets involved
in the transport process at these temperatures. This conclusion is further
endorsed by the observation~\cite{Dover,Brabers} that {\it MR tracks
magnetization as a function of H at 104 K, unlike the situation at 4.2 K}.
In fact, the expected negative sign of MR starts appearing as the
temperature is increased beyond 30 K, thereby indicating that the
(ferro)magnetic layer dominates above this temperature in this compound.
It is also important to note that Sm layer also orders
ferromagnetically~\cite{Szytula} below 100 K and if this layer dominates
low temperature ($<$ 30 K) MR, one should have seen negative MR tracking
isothermal M, in sharp contrast to the experimental observations. This
establishes that it is not even the rare-earth layer, but the Ge layer,
that dominates  low temperature magnetotransport  phenomena.

Summarising, a comparison  of  MR behaviour of the compounds
LaMn$_2$Ge$_2$ and SmMn$_2$Ge$_2$ suggests that  the dominance of magnetic
layers to the magnetotransport process diminishes with decreasing
temperature, meaning thereby that there is an unusual temperature
dependence associated with the relative involvement of layers of atoms for
this process. Thus there appears to be an interesting "electronic
separation" {\it at the unit-cell} level as the temperature is lowered as
far as the magnetotransport is concerned. The above suggestion is made
under the assumption that there are no unusual band structure effect on MR
(which, if present, is by itself  again interesting); the presently
available band structure data~\cite{Szytula} do not reveal any unusual
structure at the Fermi level within the energy scale of the magnitude of
applied magnetic field and hence we advance the present line of thoughts.
At this point, it is worth mentioning that, at the time of finalising this
article,  the idea of an electronic phase separation at a length scale
much larger than the unit-cell dimensions has been proposed to explain the
colossal magnetoresistance in mixed-valent manganites~\cite{Uehara}.   It
is fascinating that some kind of  electronic separation (as far as
magnetoresistance is concerned) may be possible with the variation of
temperature {\it even at the unit-cell level} as indicated by the present
data. Realisation of this possibility will go a long way in understanding
the transport process in modern condensed matter physics.

\end{document}